\documentclass{aa} 

\usepackage{graphicx}
\usepackage{txfonts}
\usepackage{natbib}
\begin{document}
\def\gap{\;\rlap{\lower 2.5pt
 \hbox{$\sim$}}\raise 1.5pt\hbox{$>$}\;}
\def\lap{\;\rlap{\lower 2.5pt
   \hbox{$\sim$}}\raise 1.5pt\hbox{$<$}\;}
\def\bge{\begin{equation}}
\def\ede{\end{equation}}
   \title{Globular Cluster System erosion in elliptical galaxies}


   \author{R. Capuzzo-Dolcetta
          \inst{1}
          \and
          A. Mastrobuono-Battisti\inst{1}
          }

   \institute{Dep. of Physics, Sapienza, University of Roma,
              P.le A. Moro 5,
I-00185, Roma, Italy\\
              \email{roberto.capuzzodolcetta@uniroma1.it}
             \email{alessandra.mastrobuonobattisti@uniroma1.it}
             \thanks{}
             }

   \date{Received ...; accepted ...}

 
  \abstract
   {In this paper we analyze data  of 8 elliptical 
galaxies in order to study the difference between their globular cluster systems (GCSs) radial 
distributions and those of the galactic stellar component. In all the 
galaxies studied here the globular cluster system density profile is significantly flatter toward the 
galactic centre than that of stars. }
   {A flatter profile of the radial distribution of globular cluster system
respect to that of the galactic stellar component is a difference which has an 
astrophysical relevance.
A quantitative comparative analysis of the profiles may give light on both
galaxy and globular cluster formation and evolution. If the difference is 
due to erosion of the globular cluster system, the missing GCs 
in the galactic central region may have actually merged around the galactic centre 
and formed, or at least increased in mass, the galactic nucleus. 
An observational support to this is the clear correlation between the 
galaxy integrated magnitude and the number of 
globular clusters lost and that between the central massive black hole mass 
and the total mass of globular clusters lost.}
   {We fitted the stellar and globular cluster system radial 
profiles in a set of galaxies observed at high resolution. 
We saw that the globular cluster system profile is less peaked to the 
galactic centre than the stellar one is. Assuming this difference as due to 
GCS evolution starting from a radial distribution which was initially
indistinguishable from that of stars, we may evaluate the number (and mass)
of GCs `disappeared' by a simple normalization procedure.}
   {The number of missing globular clusters is significant, ranging 
from 21\% to 71\% of their initial population abundance in the eight galaxies examined. The corresponding mass 
lost to the central galactic region is $7\times 10^7$-$1.85\times 10^9\,M_\odot$. 
All this mass carried toward central galactic regions have likely had 
an important feedback on the innermost galactic region, including
its violent transient activity (AGN) and local massive black hole formation 
and growth.}
   {}


   \maketitle
%

\section{Introduction}

Many elliptical galaxies contain more or less popolous  globular cluster 
systems (hereafter GCSs), that are, usually, less concentrated towards 
the galactic centre than the bulge-halo stars. A huge amount of literature 
dedicated to GCS identification in external galaxies and study of their 
properties, since the seminal review by Harris \& Racine (1979).
Regard to the GCS radial profiles, starting from the mentioned review, we 
just remind, for the sake of example, recent papers of Bassino et al. 
(2006),  Goudfrooij et al. (2007), Lee et al. (2008), Rhode et al. 
(2007), Sikkema et al. (2006), Spitler, L.R. et al. (2006), 
Spitler, L.R. et al. (2008),  Rhode \& Zepf (2003), Peng et al. (2004)
up to Harris et al. (2009). 

Though it cannot be yet concluded that this characteristic is common to all 
galaxies, we may say no case, at present, is known where the halo stars are 
less concentrated than the GCS. Moreover, there is a general agreement 
that the difference between the two radial distributions is real and 
not caused by a selective bias.

Consequently, different hypotheses have been advanced with the purpose 
of explaining this feature. Among these, two seem the most probable:

(i) the difference between the two distributions reflects different 
formation ages of the two systems, as suggested by Harris \& Racine 
(1979) and Racine (1991); in their opinion globular clusters originated 
earlier, when the density distribution was less peaked. However, this 
hypothesis cannot explain why the two distributions are very similar 
in the outer galactic regions;

(ii) another explanation is based on the simple assumption of coeval birth of 
globular clusters and halo stars, with a further evolution 
of the GCS radial distribution, while the collisionless halo stands 
almost unchanged.

The GCS evolution is caused by both dynamical friction, which brings massive 
clusters very close to galactic centre, and tidal interaction with a compact nucleus 
(see for example Capuzzo-Dolcetta 1993). The combined effect of these dynamical 
mechanisms acts to deplete the GCSs in the  central, denser, galactic regions, 
leaving the outer profile almost unaltered, then remaining similar to the 
profile of the halo stars. The efficiency of the mentioned phenomena is higher in 
galactic triaxial potentials (see for example Ostriker, 
Binney \& Saha 1989; Pesce, Capuzzo-Dolcetta \& Vietri 1992), in 
which there is a family of orbits, the `box' orbits, which do not conserve 
any component of the angular momentum and then well populate the central 
galactic regions. Pesce et al. (1992) showed that globular clusters moving 
in box orbits lose their orbital energy at a rate an order of magnitude 
larger than those moving in loop orbits of comparable size and energy. 
These results showed the previous evaluations of the dynamical 
friction efficiency, based on very simplified hypotheses on the 
globular cluster orbit distribution that led to an overestimate of 
the dynamical braking time-scales. On the contrary, it has been 
ascertained that massive globular clusters in triaxial potentials 
are strongly braked during their motion and thus reach the inner regions 
in a relatively short time (Capuzzo-Dolcetta 1993). There, they could 
have started a process of merging, giving origin to a central massive 
nucleus (eventually a black hole) or fed a pre-existent massive object 
(Ostriker et al. 1989; Capuzzo-Dolcetta 1993). Under the hypothesis 
that the initial GCS and halo-bulge radial distributions were the same, 
an accurate analysis of the observations would allow an estimate of the 
number of `missing clusters' and therefore of the mass removed from the GCSs.

Actually, McLaughlin (1995), Capuzzo-Dolcetta \& Vignola (1997), 
Capuzzo-Dolcetta \& Tesseri (1999) and Capuzzo-Dolcetta \& Donnarumma (2001), 
scaling the radial surface profiles 
of the halo stars of a galaxy to that of its GCS, 
estimated the number of missing globular clusters as the integral of 
the difference between the two radial profiles. 
Capuzzo-Dolcetta \& Vignola (1997) and Capuzzo-Dolcetta \& Tesseri (1999) 
suggested that the compact nuclei in our galaxy, M31 and M87, as well 
as those in many other galaxies, could have reasonably sucked in a lot 
of decayed globular clusters in the first few Gyrs of life.\\

In this paper we enlarge the discussion of the comparison of GCS and 
star distribution in galaxies, dealing with eight galaxies for which 
good photometric data are available in the literature such to draw 
reliable radial profiles.

In Sect. \ref{erosion} we resume the way, discussed deeply in previously mentioned
papers, to get from the compared 
GCS-stars radial profiles in a galaxy the number and mass of GCs disappeared due 
to time evolution of the GCS; in Sect. \ref{data} we present and discuss the 
observational data, as well as the analytical fit expressions to the 
density profiles;  in Sect. \ref{correlation} we present the extension to the data of 
this work the correlation found by Capuzzo--Doletta \& Donnarumma (2001) between the mass lost by the GCS, 
the host galaxy luminosity and the mass of the galactic central massive black hole. 
Finally, in Sect. \ref{conclusions} we summarize results and draw general conclusions. 
An error analysis of the methods used is presented in Appendix.

\section{The estimate of the number and mass of globular clusters lost}
\label{erosion}

Under the hypothesis that the flattening of the GCS 
distribution in the central region compared to the distribution of the stars in the 
galactic bulge, is due to an evolution of the 
GCS, the number of GC lost to the centre of the galaxy is 
obtained by the simple difference of the (normalized) density profiles
integrated over the whole radial range (see McLaughlin 1995). 
A general discussion of the problems in evaluating the number and mass of missing 
clusters in this way
can be found in Capuzzo-Dolcetta \& Vignola
(1997), Capuzzo-Dolcetta \& Tesseri (1999) and Capuzzo-Dolcetta \& Donnarumma (2001)  
and so it is not worth repeated here.\\
In this paper, we assume as reliable fitting function for the GCS projected radial 
distribution
a ``modified core model'', i.e. a law
\bge \label{core}
\Sigma_{GC} (r)= \frac{\Sigma_0}{\left[1+(r/r_c)^2 \right]^\gamma}
\ede
where $\Sigma_0$, $r_c$ and $\gamma$ are free parameters. The choice of this function
to fit the GCS profiles is motivated by the good agreement found with almost all the
data used for the purposes of this paper, as confirmed
by, both, the local maximum deviation of the fitting formula from the
observed data and the computed $\chi^2$. 
For the galaxy stellar profile, $\Sigma_s(r)$, we rely on the fitting formulas
provided by the authors of the various papers where we got the data from, 
that may change case by case, checking how good are the approximations to observed data.\\
The ``initial'' distribution of GCs, $\Sigma_{GC,0}(r)$, (assumed to be equal in shape 
to the present stellar profile), can be practically obtained by a vertical translation of 
the stellar profile, $\Sigma_s(r)$,  to the present  GCS distribution, $\Sigma_{GC}(r)$ 
 as given by Eq. \ref{core}.  We calculate the number of missing (lost) clusters 
as the surface integral of the difference between $\Sigma_{GC,0}(r)$ and $\Sigma_{GC}(r)$ 
over the radial range $[0, r_{max}]$ of  difference of these two profiles:
\bge \label{lostGCs}
N_l=2\pi \int_{0}^{r_{max}}\left(\Sigma_{GC,0}(r)-\Sigma_{GC}(r)\right)rdr.
\ede
The present number of GCs, $N$, is obtained integrating $\Sigma_{GC}(r)$ over the 
radial range, $[r_{min}, R]$, covered by the observations (for the value of $R$ we rely 
on the papers where we got the GCs distribution data from). The values of $N$ obtained 
with this method are usually different from those given by the authors of the papers, but 
for the purposes of this paper what is important is the difference bewteen $N$ and $N_l$ and, 
so, it is crucial a homogeneous way to determine them.\\
The initial number of GCs in a galaxy is, indeed, estimated as $N_i=N_l+N$.\\
The numerical values of $N_l$, $N_i$ and $N$  are functions of the fitting parameters and 
of the integration limits. 
These dependences and their contribution to the errors on the final results are discussed 
in Appendix. \\
An estimate of the mass removed from the GCS ($M_l$) can be obtained through the number of 
GCs lost, $N_l$, and the estimate of the mean mass of the missing globular clusters, 
$\left\langle m_l\right\rangle$. A priori,
the determination of $\left\langle m_l\right\rangle$ needs the knowledge of the 
initial mass spectrum of the CGS, which has suffered of an evolutionary erosion. 
However, the most relevant evolutionary phenomena (tidal shocking and dynamical friction) act 
on opposite sides of the initial mass
function, and so we expect that the mean value of the globular cluster mass has
not changed very much in time (see Capuzzo-Dolcetta \& Tesseri,
1997 and Capuzzo-Dolcetta \& Donnarumma 2001). Hence, we can assume the present mean 
value of the mass of
globular clusters, $\left\langle m\right\rangle$, as a good reference value for 
$\left\langle m_l\right\rangle$.\\
For NGC 4374, NGC 4406 and NGC 4636 we calculated $\left\langle m \right\rangle$ 
using their GC luminosity functions (GCLFs) and assuming the same typical 
mass-to-light ratio of GCs in our Galaxy, i.e. $(M/L)_{V,\odot}=1.5$ for NGC 4406 and 
NGC 4636 or $(M/L)_{B,\odot}=1.9$ \cite{Ill} in the case of NGC 4374. For NGC 4636 
we used also the mass function that represents its present distribution 
of GCs (see Sect. \ref{NGC 4636}). For the remaining galaxies (NGC 1400, NGC 1407, 
NGC 4472, NGC 3258 and NGC 3268) there is no better way to estimate  the total 
mass of `lost' GCs than adopting as a `fiducial' reference value for their mean 
mass, $\left\langle m_l\right\rangle$, the value, 
$\left\langle m_{MW}\right\rangle=3.3\times 10^5M_\odot$, of the present mean 
GC mass in our Galaxy.\\


\begin{figure*}
\centering
\includegraphics[width=17cm]{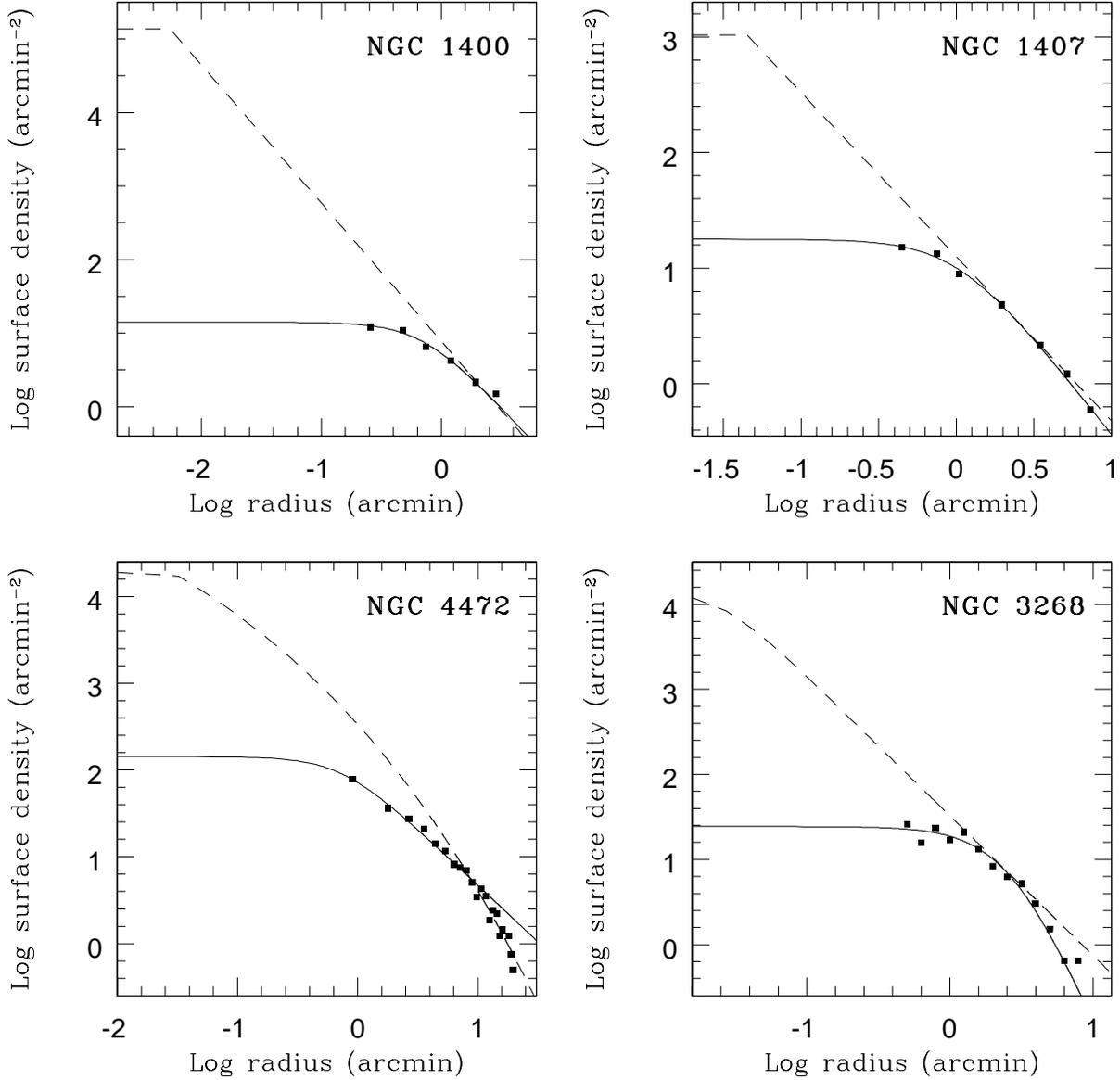}
\caption{Surface number density for NGC 1400, NGC 1407, NGC 4472 (M 49) 
and NGC 3268. Black squares represent the observed GC distribution; 
the solid line is its modified core  model fit. The 
dashed curve is the surface brightness profile of the underlying 
galaxy (a power law and a central flat core for NGC 1400 and NGC 1407, a Sersic core model 
for NGC 4472 and a Nuker law for NGC 3268), vertically normalized 
to match the radial profile of the GCS in the outer regions.}
\label{fig1}
\end{figure*}

\begin{figure*}
\begin{center}
\includegraphics[width=17cm]{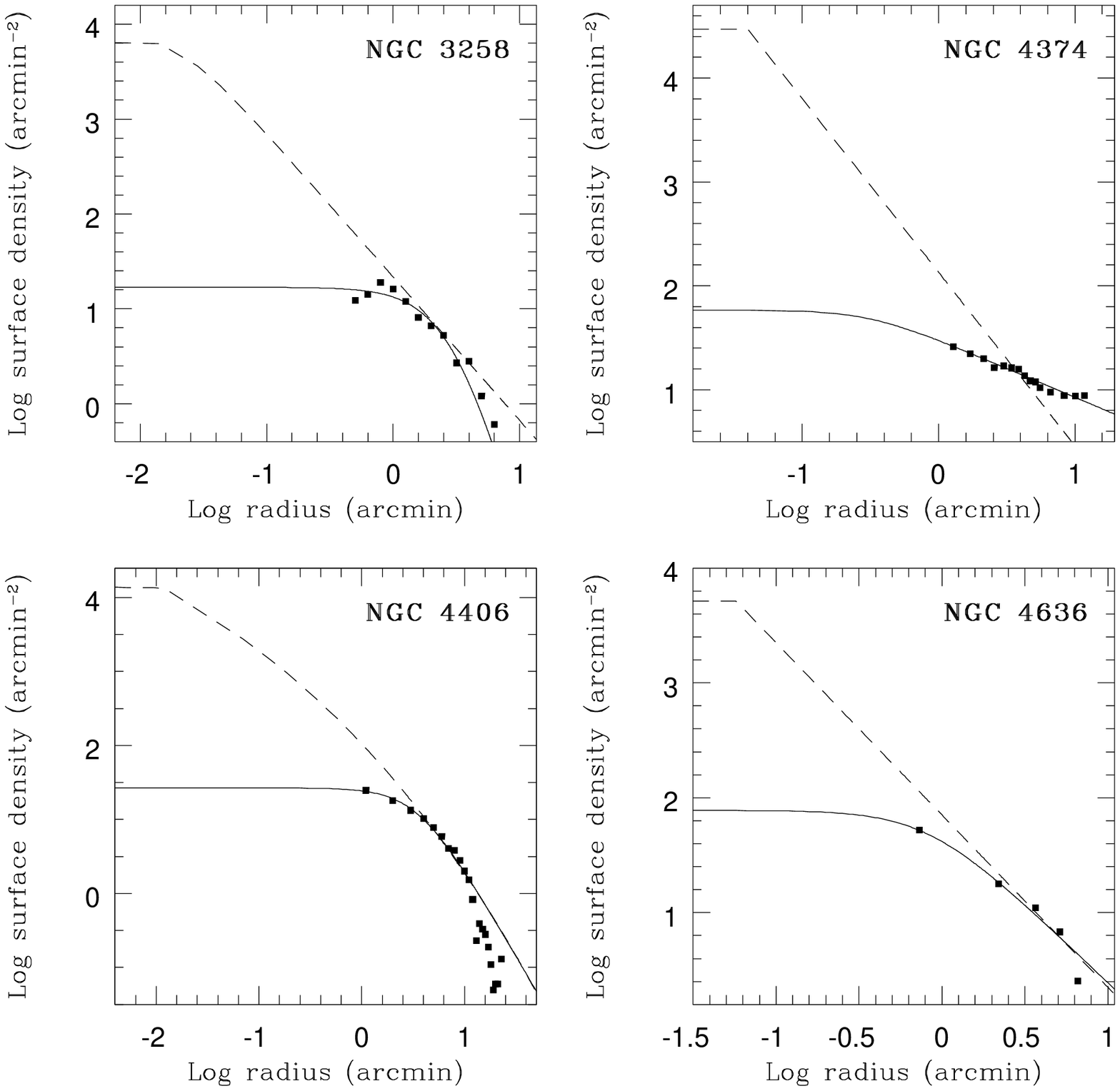}
\caption{Surface number density for NGC 3258, NGC 4374, NGC 4406 and NGC 
4636. Black squares represent the observed GC distributions; 
the solid lines are their modified core  model fit. The dashed curves are 
the surface brightness profile of the underlying galaxy (a Nuker law for 
NGC 3268, a power law and a central flat core for NGC 4374 and NGC 4636 and a Sersic 
core model for NGC 4406), vertically normalized to match the radial profile of the
cluster system in the outer regions.}\label{fig2}
\end{center}
\end{figure*}


\section{Data and results}\label{data}  
The data for the study of this paper have been collected from the literature. We will 
analyze a set of 8 galaxies for which the GC content and the radial profile has been 
reliably determined and apt to our purposes. \\ 
The galaxies are: NGC 1400,  NGC 1407,  NGC 4472 (M 49),  NGC 3268,  NGC 3258, NGC 4374, 
NGC 4406 and NGC 4636. These galaxies go to enlarge the set of seventeen galaxies (Milky Way, M 31, M 87, 
NGC 1379, NGC 1399, NGC 1404, NGC 1427, NGC 1439, NGC 1700, NGC 4365, NGC 4494, NGC 4589, 
NGC 5322, NGC 5813, NGC 5982, NGC 7626, IC 1459) whose GCS radial profiles have been 
compared to the stellar distribution in previous papers (\citep{McL95}, 
Capuzzo-Dolcetta \& Vignola 1997, 
Capuzzo-Dolcetta \& Tesseri 1999, Capuzzo-Dolcetta \& Donnarumma 2001).
\subsection{NGC 1400} \label{NGC 1400}
The surface density profile of this galaxy is given by Forbes et al. (2006)
(hereafter F06) who fitted it by mean of a power law $\Sigma_s(r) \propto r^{-1.88}$. 
This fitting law is reliable outside the galactic core, i.e. for $r>r_b=0.0055$\,arcmin 
(Spolaor et al. 2008).  The luminosity profile of the galaxy for $r\leq r_b$ is almost flat, 
and linked to the external power law. 
We fitted the GCS distribution by a modified core model with $\Sigma_0 = 14.1$ 
arcmin$^{-2}$,
$r_c=0.7$ arcmin and $\gamma=0.88$.
Integrating $\Sigma_{GC}(r)$ in the radial range where GCs are observed, i.e. from 
$r_{min}=0$\,arcmin to $R=2.8$\,arcmin, we obtain $N=73$ as the present number of 
GC in NGC 1400.
The initial GCS distribution results to be approximated by: 
\begin{eqnarray}
\Sigma_{GC,0}(r) = \left\{ \begin{array}{ll}
1.37\times 10^5\;\textrm{arcmin}^{-2} & \textrm{$r\leq r_b$}\\
7.76 r^{-1.88} & \textrm{$r>r_b$}\\
\end{array} \right.
\end{eqnarray}
(see Fig. \ref{fig1}).\\
Using the general method described in Sec. \ref{erosion} and the estimated value 
$r_{max}=2.3$\,arcmin we have that the number of missing clusters in this galaxy is 
$N_l=183$, i.e. about 71\% of the initial population of globular clusters, 
$N_i=N_l+N=256$.\\
An estimate of the mass lost by the GCS is 
$M_l=N_l \left\langle m_{MW}\right\rangle =6.04\times 10^7M_{\odot}$. 


\subsection{NGC 1407} \label{NGC 1407}
As for NGC 1400, data for this galaxy and its GCS are taken from F06.
The luminosity profile of the galaxy stars is fitted by a power law,
$\Sigma_s(r)\propto r^{-1.42}$.
This power law fit fails in the inner region where luminosity shows a core of radius 
$r_b\simeq 0.045$\,arcmin (Spolaor et al. 2008). As for NGC 1400 we thus assume, 
for $r\leq r_b$, a flat distribution matched to the external power law.
The GCS modified core model has, in this case,
$\Sigma_0= 17.8$\,arcmin$^{-2}$, $r_c =1.02$\,arcmin and $\gamma = 0.85$ as 
better fitting parameters.
The normalizing vertical translation of the stellar profile leads to
\begin{eqnarray}
\Sigma_{GC,0}(r) = \left\{ \begin{array}{ll}
1.04\times 10^3\;\textrm{arcmin}^{-2} & \textrm{$r\leq r_b$}\\
12.6r^{-1.42} & \textrm{$r>r_b$}\\
\end{array} \right.
\end{eqnarray}
(see Fig. \ref{fig1}).\\
Integrating the difference of the GCS ``initial'' and present radial profiles in 
the galactic region where these differ, i.e. up to $r_{max}=2.34$\,arcmin 
(see Eq. \ref{lostGCs}), we obtain $N_l=84$. The present  number of GC obtained 
integrating $\Sigma_{GC}(r)$ in the radial range covered by the observations 
(i.e. from 0 to $R=7.3$\,arcmin) is $N=314$. 
The GCS has therefore lost 21\% of its initial population, $N_i=398$.\\
Also in this case, we can evaluate the mass lost by the system as $M_l=N_l \left\langle m_{MW}\right\rangle =2.77\times 10^7M_{\odot}$.

\subsection{NGC 4472 (M 49)} \label{M 49}
Data for the GC distribution in this giant elliptical galaxy in Virgo are taken 
from C\^ot\'e et al. (2003).
The galaxy star luminosity distribution, according to Ferrarese et al (2006),
is well reproduced by a Sersic core model (Trujillo et al. 2004), i.e. by
\bge \label{Sersic}
\Sigma_s(r)=\Sigma_b\left\{ {\left(\frac{r_b}{r}\right)}^\gamma \theta (r_b-r)+\textrm{e}^{b_n\left(\frac{r_b}{r_e}\right)^{\frac{1}{n}}}\theta(r-r_b)
\textrm{e}^{-b_n\left(\frac{r}{r_e}\right)^{\frac{1}{n}}}\right\},
\ede
where $\Sigma_s(r_b)=\Sigma_b$, $\theta (x)$ is the usual Heaviside function,
$r_b$ ({\it break} radius) divides the profile into an inner ($r\leq r_b$) power-law region and an outer ($r \geq r_b$) exponential region; $r_e$ is the `effective' radius and
$b_n =1.992n-0.3271$, with $n$ free fitting parameter. For M 49 the parameter values are (C\^ot\'e et al. 2003):
$\gamma=0.086$, $n=5.503$, $b_n=10.635$, $r_e=208$ arcsec ($=16.89$ kpc), $r_b=1.94$ arcsec ($=0.158$ kpc).\\
In this case the core model fit to the observed GCS distribution gives $\Sigma_0= 6$\nolinebreak[4]\,kpc$^{-2}$, $r_c =3.57$\,kpc and $\gamma = 0.652$ as optimal parameters.
The usual vertical translation leads to 
\begin{eqnarray} 
\Sigma_{GC,0}(r)=722 \left\{ {\left(\frac{0.158}{r}\right)}^{0.086}\theta (0.158-r)+\right.\nonumber\\ \left. +94.7\theta(r-0.158)
\textrm{e}^{-10.635\left(\frac{r}{16.89}\right)^{0.18}}\right\}.
\end{eqnarray}
Fig. \ref{fig1} shows how the modified core model profile does not represent very well the
observed GCS external profile, leading, there, to overestimate.\\
The surface integral (Eq. \ref{lostGCs}), performed with $r_{max}=42.34$\,kpc, 
gives $N_l=5598$. Integrating $\Sigma_{GC}(r)$ up to $R=100$\,kpc we have that $N=6334$. 
Hence, M 49 has lost 47\% of the initial population of its GCs, $N_i=11,932$.\\
As for NGC 1400 and NGC 1407, no better estimate of $\left\langle m_l\right\rangle$ is available, so we evaluate $M_l$ as 
$M_l=N_l\left\langle m_{MW}\right\rangle =1.85\times10^9M_{\odot}$.

\subsection{NGC 3268} \label{NGC 3268}
The GC distribution of this galaxy is discussed by Dirsch, Richtler \& Bassino (2003). The resulting
core model fit parameters have the values: $\Sigma_0=24.4$ arcmin$^{-2}$,
$r_c=2.6$\,arcmin and $\gamma=1.9$. The stellar luminosity profile, following Capetti \& Balmaverde (2006), is well represented by  
a ``Nuker'' law (introduced by Lauer et al. 1995):
\bge \label{Nuker}
\Sigma_s(r)=2^{(\beta-\gamma)/\alpha}\Sigma_b \left( \frac{r_b}{r}\right)^\gamma 
\left[1+\left(\frac{r}{r_b}\right)^\alpha\right]^{(\gamma-\beta)/\alpha}    
\ede
where $\beta$ is the slope of the external region of luminosity profile, $r_b$ is the ``break radius'' (corresponding to a brightness $\Sigma_b$) where the profile flattens to a smaller slope, measured 
by the parameter $\gamma$; $\alpha$ sets the sharpness of the transition between the inner and outer profile.
In the case of NGC 3268: $\alpha=2.49$, $\beta=1.64$, $\gamma=0.13$ and $r_b=0.0252$\,arcmin.
By the usual vertical translation of this profile, the initial GCS profile is obtained
\begin{equation}
\label{eq:Sigma}
\Sigma_{GC,0}(r)=13692 \left( \frac{0.0252}{r}\right)^{0.13} 
\left[1+\left(\frac{r}{0.0252}\right)^{2.49}\right]^{-0.606}.    
\end{equation}
The departure, visible in Fig. \ref{fig1}, of the profile given by Eq.\ref{eq:Sigma} from the modified core model profile
in the external galactic region is mainly due to the incompleteness of the GC detection in the outermost regions.
Integrating $\Sigma_{GC}(r)$ from $r_{min}=0$\,arcmin to $R=7.94$\,arcmin we have $N=505$ as present GC number.\\
The number of globular clusters lost is found to be $N_l=398$ (from 0 to $r_{max}=2.3$\,arcmin), that means about 44\% of the initial 
abundance, $N_i=913$.\\
Also in this case, to evaluate the mean mass of lost GC in NCG 3268 we have to assume 
$\left\langle m_l \right\rangle=\left\langle m_{MW}\right\rangle$, 
obtaining $M_l=N_l \left\langle m_{MW}\right\rangle=1.31\times 10^8M_{\odot}$.

\begin{table*}
\small
\begin{center}
\begin{tabular}{l c c c c c c} 
\hline
{Galaxy} & {$\Sigma_0$}& {$r_c$} & {$\gamma$} & {$r_{max}$}& {$R$}
\\
\hline

\textbf{NGC 1400}  & 14.1 & 0.7 & 0.88 &2.3& 2.8\\

\textbf{NGC 1407}  & 17.8 & 1.02 & 0.85 &2.34 & 7.3\\

\textbf{NGC 4472} &142& 0.73 & 0.652 & 8.69& 20.52\\

\textbf{NGC 3268} & 24.4 & 2.6 & 1.9 & 2.3 & 7.94\\

\textbf{NGC 3258} & 16.9 & 3.1 & 2.4 &  2.16 & 7.94\\

\textbf{NGC 4374} & 58.4 & 0.31 & 0.278 &  3.84 & 11.8\\

\textbf{NGC 4406} & 26.76 & 3.52 & 1.19 & 5.36 & 24\\

\textbf{NGC 4636} & 77.66 & 0.823 & 0.691 & 4.75 & 6.6\\

\hline

\end{tabular}
\caption {Col. 1: galaxy name; col. 2, 3 and 4: parameters  
of the modified core model fit for all the galaxies studied; 
col. 5: upper limit  in the integral giving the number of the lost GCs ($r_{max}$); 
col. 6: upper limit  of the integral performed to estimate the present number of 
GCs ($R$). $\Sigma_0$ is in arcmin$^{-2}$; $r_c$, $r_{max}$ and $R$ are in arcmin.} \label{tab1}
\end{center}
\end{table*}

\begin{table*}[!t]
\footnotesize
\begin{center}
\begin{tabular}{l c c c c c c c c c} 
\hline
{Galaxy}&{Model}& {$\eta$} & {$r_b$} & {$r_e$} & {$b_n$}& {$n$} & {$\alpha$} & {$\beta$} &{$\gamma$}
\\
\hline

\textbf{NGC 1400} &lc& $7.76$ & $5.5\times 10^{-3}$&-&-&-&  $1.88$ &-&-\\

\textbf{NGC 1407} &lc& $12.6$ &0.045&-&-&-&  $1.42$&-&- \\

\textbf{NGC 4472} &cS& $17139$ & 0.0323 & 3.47 & 10.635 & 5.503&-&-& 0.086 \\

\textbf{NGC 3268} &N&13692&0.0252&-&-& -&2.49&1.64&0.13\\

\textbf{NGC 3258} &N& 8489 & 0.0192 &-&-& -& 2.10 & 1.51 & 0 \\

\textbf{NGC 4374} &lc& $135$& 0.0398&-&-&-& $1.67$&-&-\\

\textbf{NGC 4406} &cS& 13490& 0.012 & 6.86 & 13.649 & 7.02 &-&-& 0.021 \\

\textbf{NGC 4636} &lc& $70.8$  &0.0573&-&-&-&  $1.5$&-&-\\
\hline

\end{tabular}
\caption{Galactic luminosity fitting parameters. Col. 1: galaxy name; col. (2) 
key identifying the galaxy light profile model (lc=linear with a flat core in the inner region, 
cS=core-S\'{e}rsic, N=Nuker); col. 3-10: parameters of the various profile models 
(see Sect. \ref{erosion} and \ref{data} for details). $\eta$ is in arcmin$^{-2}$; $r_b$ and $r_e$
are in arcmin.} \label{tab2}
\end{center}
\end{table*}

\begin{table*}[h]
\small
\begin{center}
\begin{tabular}{l r r r r r r r} 
\hline 
{Galaxy} & {$N$} & {$N_i$} & {$N_l$} & {$\delta N$} & {$\epsilon_l$} &{$M_i$}& {$M_l$} 
\\
\hline

\textbf{NGC 1400} & 73 & 256 & 183 & 0.71 & 0.40 & $8.45\times 10^7$ &  $6.04\times 10^7$\\

\textbf{NGC 1407} & 314 & 398  & 84 & 0.21 & 0.12& $1.31\times 10^8$ & $2.77\times 10^7$ \\

\textbf{NGC 4472} & 6334 & 11932 & 5598 & 0.47 & 0.20& $3.94\times 10^9$ & $1.85\times 10^9$  \\

\textbf{NGC 3268} & 505 & 903 & 398 & 0.44 & 0.15 & $2.98\times 10^8$ & $1.31\times 10^8$ \\

\textbf{NGC 3258} & 343 & 555 & 212 & 0.38 & 0.16 & $1.83\times 10^8$ & $7.00\times 10^7$ \\

\textbf{NGC 4374} & 4655 & 7016 & 2361 & 0.34 & 0.050 & $2.34\times 10^9$ & $7.86\times 10^8$ \\

\textbf{NGC 4406} & 2850 & 4209 & 1359 & 0.32 & 0.23 & $1.25\times 10^9$ & $4.04\times 10^8$ \\

\textbf{NGC 4636} & 1411 & 2157 & 746 & 0.35 & 0.11 & $6.41\times 10^8$ & $2.22\times 10^8$ \\

\hline
\end{tabular}
\caption{col. (1): galaxy name; col. 2-8: the present number of GCs ($N$), 
its initial value ($N_i$), the number of GCs lost ($N_l$), 
the percentage of GCs  lost and the estimated relative error on $N_l$ ($\epsilon_l$, 
see Appendix), the estimate of the initial mass of the whole GCS ($M_i$) 
and of the mass lost by each GCS ($M_l$).$M_i$ and $M_l$ are in solar masses.} \label{tab3}
\end{center}
\end{table*}

\subsection{NGC 3258} \label{NGC 3258}
As for NGC 3268 the GCS density profile data for NGC 3258 are taken from Dirsch, Richtler \& Bassino (2003).
The best modified core model fit is given by the values $\Sigma_0=16.9$\,arcmin$^{-2}$, $r_c=3.1$\,arcmin and $\gamma=2.4$. 
The analytical fit to the luminosity profile of the galaxy is, again, 
 obtained with the ``Nuker'' law (Eq. (\ref{Nuker})) with $\alpha=2.10$, 
$\beta=1.51$, $\gamma=0$ and $r_b=0.0192$\,arcmin. By mean of the usual procedure, 
the initial GCS profile is obtained: 
\bge
\Sigma_{GC,0}(r)=8489 \left[1+\left(\frac{r}{0.0192}\right)^{2.10}\right]^{-0.719}.
\ede
The present number of GCs is $N=343$ (with $R=7.94$\,arcmin). Performing the 
surface integral of the difference of the initial and present distribution in 
the radial range up to $r_{max}=2.16$\,arcmin 
(see Eq. \ref{lostGCs}) we have $N_l=212$, corresponding to 38\% of 
the initial GCS population, $N_i=555$.\\
For this galaxy, we obtained  $M_l= N_l \left\langle m_{MW}\right\rangle=7.00\times 10^7M_{\odot}$. 


\subsection{NGC 4374 (M 84)} \label{NGC 4374}
Gomez \& Richtler (2004) studied the GCS of this giant elliptical galaxy, using
photometry in the $B$ and $R$ bands, to draw its radial surface distribution. Also in this case,
the profile of the GC number density is flatter than the galaxy light (see Fig. \ref{fig2}).
The best modified core model fit to the GC data is given by $\Sigma_0=58.4$\,arcmin$^{-2}$,
$r_c=0.31$\,arcmin and $\gamma=0.278$. \\
The galaxy light is characterized by a central core of radius $r_b\simeq 0.0398$\,arcmin (Lauer et al., 2007); for $r>r_b$, it is well fitted  
by the power law $\Sigma_s(r)\propto r^{-1.67}$ (Gomez \& Richtler 2004). 
The usual normalization leads to
\begin{eqnarray}
\Sigma_{GC,0}(r) = \left\{ \begin{array}{ll}
2.94\times 10^4\;\textrm{arcmin}^{-2} & \textrm{$r\leq r_b$}\\
135r^{-1.67} & \textrm{$r>r_b$},\\
\end{array} \right.
\end{eqnarray}
as GCS initial radial profile.
Integrating our core model up to $R=11.8$\,arcmin we get $N=4655$ as present number of GCs. The usual integration of the difference of the initial and present GC distribution (Eq. \ref{lostGCs} with $r_{max}=3.84$\,arcmin) leads to $N_l=2361$.
Hence NGC 4374 has lost 34\% of its initial population of globular clusters, $N_i=7016$.\\
In the case of NGC 4374 the value of the mean mass of a GC has been evaluated using the GCLF in the R band given by Gomez \& Richtler (2004). 
The mean color $\left\langle (B-R)_0\right\rangle=1.18$ of GCs in this galaxy 
(Gomez \& Richtler 2004) allows us to estimate the mean B absolute magnitude and the 
mean luminosity of GCs in the B band, $\left\langle (L/L_B)_\odot\right\rangle$ assuming $m-M= 31.61$ (Gomez \& Richtler 2004). It results $\left\langle (L/L_B)_\odot\right\rangle=1.75\times 10^5$ 
(with $M_{B,\odot}=5.47$, Cox 2000).\\
Adopting the mass to light ratio  $\left(M/L)_{B,\odot}\right)=1.9$ 
obtained by Illingworth (1976) for 10 galactic globular clusters, 
we get $\left\langle  m_l \right\rangle=3.33\times 10^5 M_{\odot}$ 
and  $M_l=N_l\left\langle m_l\right\rangle=7.86\times 10^8M_{\odot}$. 
\subsection{NGC 4406 (VCC 881)} \label{NGC 4406}
NGC 4406 is another giant elliptical in Virgo; its GCS has been studied by mean of
the Mosaic Imager on the 4m Mayall telescope at the KPNO (Rhode \& Zepf 2004) in
the $B$, $V$ and $R$ bands. The resulting best fit core model is characterized by 
$\Sigma_0=26.76$ arcmin$^{-2}$, $r_c=3.52$ arcmin and $\gamma=1.19$.
The galaxy light is well fitted by a Sersic core model (Eq. \ref{Sersic}), whose parameters have been determined
by Ferrarese et al. (2006). Its vertical translation gives the initial GCS radial profile
\begin{eqnarray}
\Sigma_{GC,0}(r) &=&13490 \left[ \left(\frac{0.012}{r}\right)^{0.021}\theta (0.012-r)\right.+\nonumber\\&+&\left.
250\theta(r-0.012)
\textrm{e}^{-13.649\left(\frac {r}{6.86}\right)^{0.142}} \right].
\end{eqnarray}
Integrating the present distribution of GCs, from $r_{min}=0$\,arcmin to $R=24$\,arcmin, we have $N=2850$.
The surface integral given in Eq. \ref{lostGCs}, with $r_{max}=5.36$\,arcmin, gives the number of globular clusters lost, $N_l=1359$, i.e. about 32\% of the initial GC population.\\ 
Using the GCLF of this galaxy and its distance modulus $m-M=31.12$ (Rhode \& Zepf, 2004),  we evaluated the mean value
of the absolute GC V magnitude, $\left\langle M_V \right\rangle= -8.42$ which corresponds to the mean luminosity
$\left\langle L/L_\odot\right\rangle_V=1.98\times 10^5$ ($M_{V,\odot}=4.82$ from Cox 2000).\\ 
Assuming $(M/L)_{V,\odot}=1.5$, we obtain $\left\langle m_l\right\rangle=2.97\times 10^5M_{\odot}$. 
This estimate leads to the value of the mass lost by the GCS, $M_l= N_l\left\langle m_l\right\rangle=4.04\times 10^8M_{\odot}$.  

\subsection{NGC 4636} \label{NGC 4636}
The GC content of this galaxy has been studied by Kissler Patig et al. (1994).
The modified core model fit has $\Sigma_0=77.66$ arcmin$^{-2}$, $r_c=0.823$ arcmin 
and $\gamma=0.691$ as optimal parameter values. \\
The galactic light profile shows an inner flat distribution (a core with radius $r_b\simeq 0.0573$\,arcmin (Lauer et al. 2007)), while for  $r>r_b$ the light distribution is well fitted by the power law fit $\Sigma_s(r) \propto r^{-1.5}$ (Kissler Patig et al. 1994).\\
The vertical translation of the stellar profile gives the initial GCS profile:
\begin{eqnarray}
\Sigma_{GC,0}(r) = \left\{ \begin{array}{ll}
5.16\times 10^3\;\textrm{arcmin}^{-2} & \textrm{$r\leq r_b$}\\
70.8r^{-1.5} & \textrm{$r>r_b$}.\\
\end{array} \right.
\end{eqnarray}
Integrating the present surface density profile of the GCS up to $R=6.6$\,arcmin, 
we obtain $N=1411$. Performing the surface integral given in Eq. \ref{lostGCs} 
(with $r_{max}=4.75$\,arcmin), we estimate that the number of GCs disappeared 
is $N_l=746$, i.e. 35\% of the initial population, $N_i=2157$.
In the case of this galaxy we obtained two different estimates of the mass lost by the GCS, starting from data taken from Kissler Patig et al. (1994). 
The first estimate has been obtained using the GCLF (Kissler Patig et al. 1994). 
As for NGC 4406 we calculated the mean absolute V magnitude of GCs, 
$\left\langle M_V\right\rangle= -8.07$, (given $m-M=31.2$ by Kissler Patig et al. 1994). 
Assuming for GCs in NGC 4636 the same $M/L_V$ ratio of galactic GCs, $(M/L_V)_{\odot}=1.5$, 
the deduced mean luminosity of GCs, $\left\langle L/L_\odot\right\rangle_V=1.43\times 10^5$, gives $\left\langle m_{l,1}\right\rangle=2.15\times 10^5M_{\odot}$, and so $M_{l,1}= N_l\left\langle m_{l,1}\right\rangle=1.29\times 10^8M_{\odot}$. \\
Another estimate is found using the mass distribution of GCs obtained in Kissler Patig et al. (1994) transforming the magnitude bins of the GCLF candidates into masses using the relation given by Mandushev et al. (1991): $\textrm{log}(M/M_{\odot})=-0.46 M_V+1.6$ (corresponding to a mean mass to light ratio $(M/L)_{V,\odot}\simeq 2.0$). Knowing the mass distribution we can directly calculate the mean mass of GCs, $\left\langle m_{l,2}\right\rangle=3.79\times 10^5M_{\odot}$, and thus $M_{l,2}= N_l\left\langle m_{l,2}\right\rangle=2.97\times 10^8M_{\odot}$. \\ 
The averages of our two estimates gives $M_l=2.22\times 10^8M_{\odot}$. 

Tables \ref{tab1}, \ref{tab2} resume the parameters of the radial profile
fitting functions for the studied galaxies, while \ref{tab3} resume the results
in terms of estimated number and mass of GC lost.

\section{The correlation between $M_l$, $M_V$ and $M_{bh}$}\label{correlation}

\begin{figure*}
   \centering
  \resizebox{\hsize}{!}{\includegraphics[clip=true]{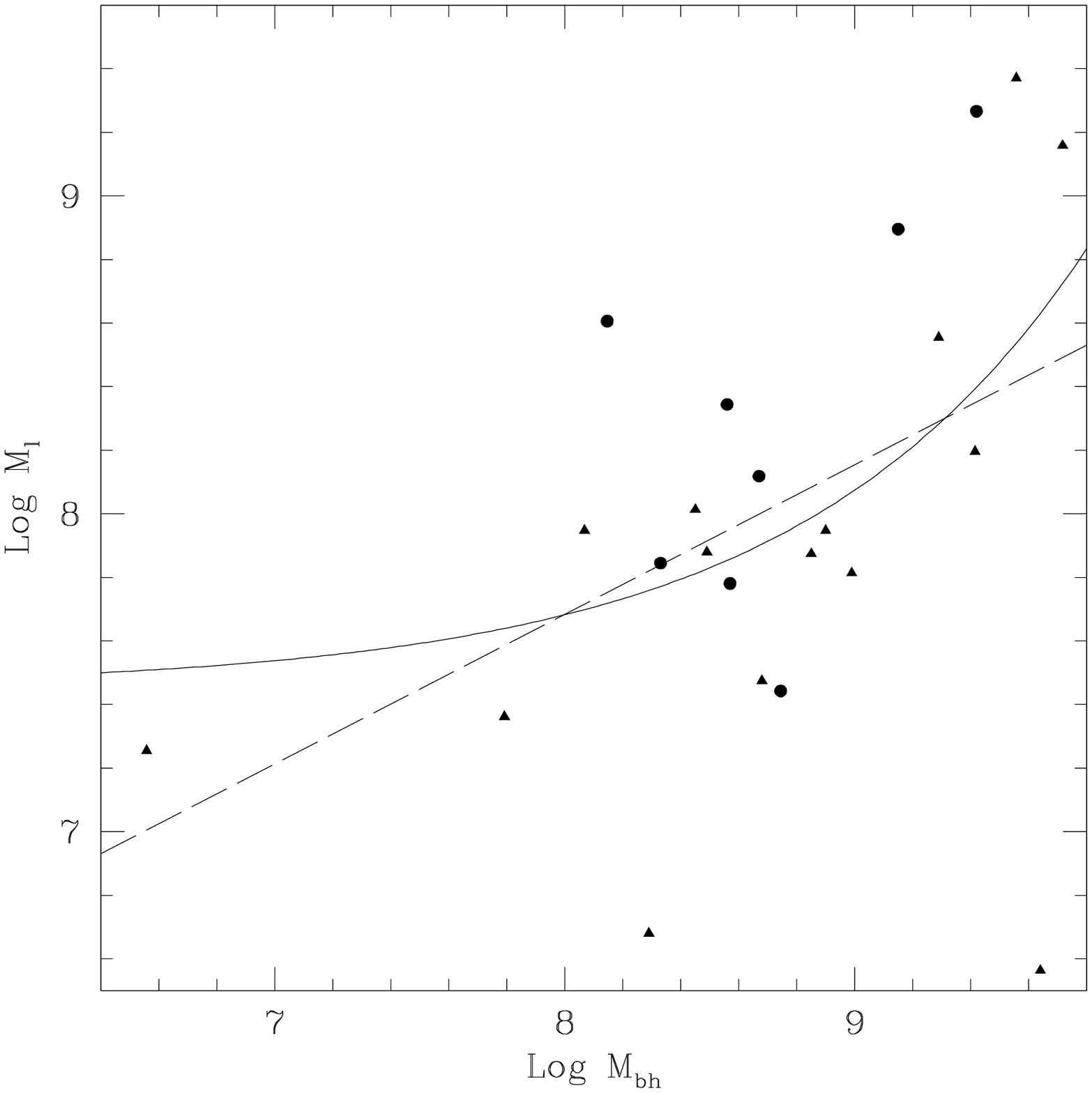}
   \includegraphics[clip=true]{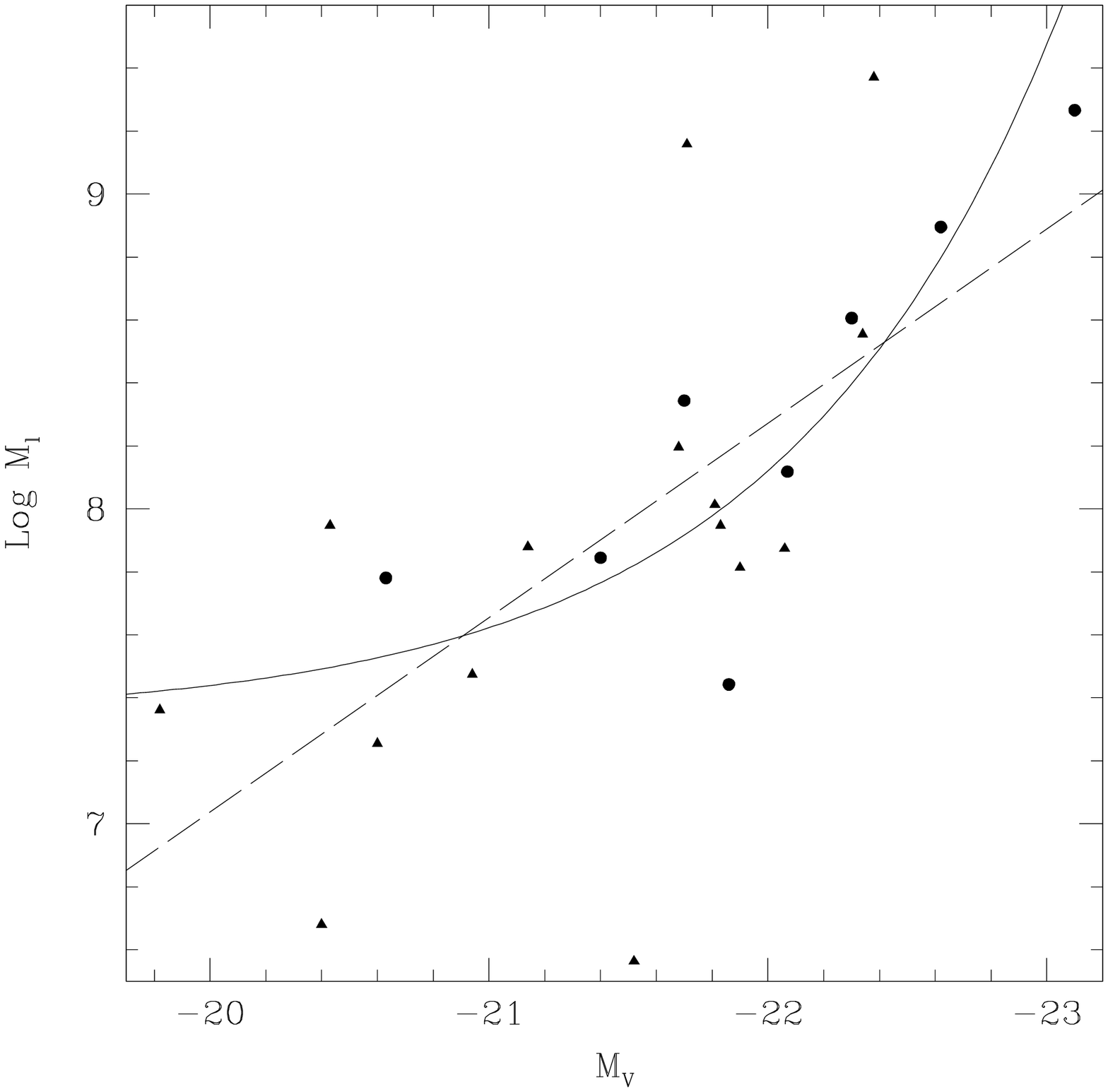}}
     \caption{The correlation between the GCS (logarithmic) mass lost and the 
central galactic black hole mass (left panel) and integrated V magnitude of the host galaxy 
(right panel) for the set of galaxies in Table \ref{tab4}. Masses are in solar masses.
Black circles represent the eight galaxies whose GCS data are discussed in this paper,
black triangles refer to the others. The straight lines and curves are the approximation fits 
discussed in Sect.\ref{correlation}.}  
        \label{fig3}
\end{figure*}

The evolutionary explanation of the difference between the initial and present GC
distribution implies a correlation between the (supposed) mass lost by GCS 
with the mass of the galactic central supermassive black hole ($M_{bh}$) and. likely,
with the host galaxy luminosity ($M_V$). Tab. \ref{tab4} reports the whole set of galaxies 
for which we have the estimate of $M_V$, $M_{bh}$ and $M_l$.\\
Fig. \ref{fig3} shows a plot of Tab. \ref{tab4} data which clearly indicate 
an increasing trend of $M_l$ as function of $M_{bh}$ (left panel) and of
$M_V$ (right panel).

In particular, the linear fit of data in the left panel is given 
by $\log M_l= a \log M_{bh} + b$ with $a\pm \sigma(a)=0.47\pm 0.20$ and 
$b\pm \sigma(b)=3.9\pm 1.8$, giving $r=0.45$ and $\chi^2=9.5$. 
The, alternative, exponential fit gives 
$\log M_l= \alpha \exp (\log M_{bh} )+ \beta$ 
where $\alpha\pm \sigma(\alpha)=(7.7\pm 3.3)\times 10^{-5}$ 
and $\beta\pm \sigma(\beta)=7.45\pm 0.28$, $r^2=0.20$ and $\chi^2=9.6$.\\ 
We performed also fits for for the correlation between $M_l$ and $M_{bh}$
excluding the data which have a great residual from the mean square 
(those of NGC 1439 and NGC 1700) obtaining the following 
parameters of the linear fit: $a\pm \sigma(a)=0.61\pm 0.15$ and 
$b\pm \sigma(b)=2.9\pm 1.3$, with $r=0.69$ and $\chi^2=4.0$. 
The exponential fit parameters are in this case: 
$\alpha\pm \sigma(\alpha)=(1.17\pm 0.22)\times 10^{-4}$ 
and $\beta\pm \sigma(\beta)=7.31\pm 0.18$, giving $r^2=0.60$ and $\chi^2=3.1$.

The least square, straight-line fit to the whole set of data shown in the right 
panel of Fig. \ref{fig3} 
is given by $\log M_l= a M_V + b$ with $a\pm \sigma(a)=-0.62\pm 0.15$ 
and $b\pm \sigma(b)=-5.3\pm 3.2$, giving $r=0.67$ and $\chi^2=6.6$. The 
exponential fit on the same data gives $\log M_l= \alpha \exp(-M_V)+ \beta$ where 
$\alpha\pm \sigma(\alpha)=(2.20\pm 0.49)\times 10^{-10}$ and 
$\beta\pm \sigma(\beta)=7.33\pm 0.19$, $r^2=0.48$ and $\chi^2=6.2$.

\begin{figure*}
   \centering
  \resizebox{\hsize}{!}{\includegraphics[clip=true]{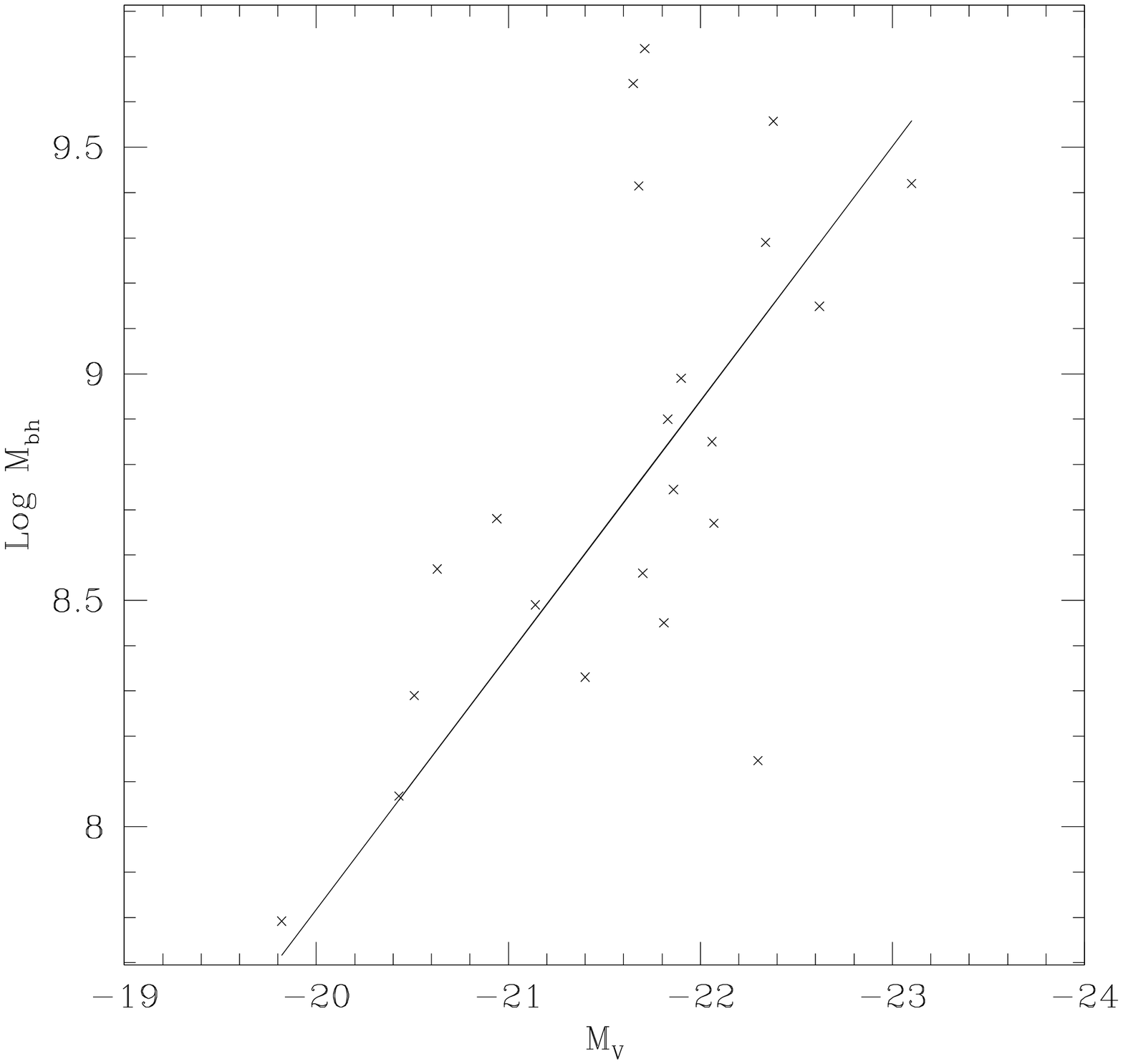}
   \includegraphics[clip=true]{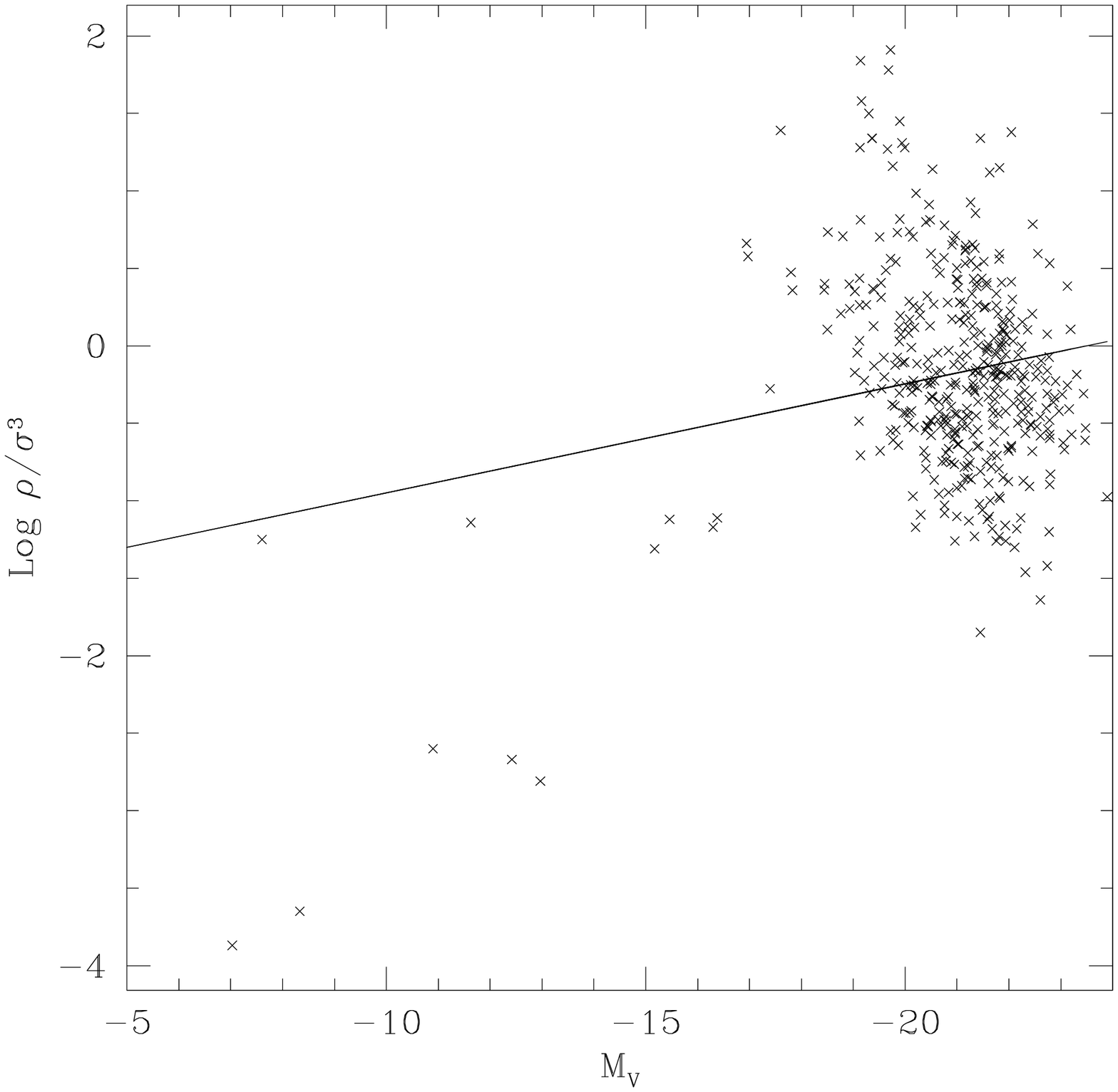}}
     \caption{Left panel: the correlation between the logarithm of the central galactic 
black hole mass (in solar masses) and the integrated V galactic magnitude (see Table \ref{tab4}). 
Right panel: the correlation between the value of the parameter proportional to the central galactic 
phase-space density (in arbitrary units) and the galactic integrated V magnitude. The straight lines are
the least-square fits to the data (see Sect. \ref{correlation}.)}  
        \label{fig4}
\end{figure*}

The correlation seen in the right panel of Fig. \ref{fig3} 
between $M_l$ and $M_V$ reflects, both, an expected physical 
dependence on the total galactic mass of evolutionary processes acting 
on GCSs and, simply, the positive correlation between $M_{bh}$ and $M_V$ for the same set 
of galaxies. Actually, the $M_{bh}$-$M_V$ correlation for the set of galaxies in Tab. 
\ref{tab4} has a clearly positive slope, as shown also by the least square fit
in Fig. \ref{fig4}. The least square fit is
$\log M_{bh}= a M_V + b$ with $a\pm \sigma(a)=-0.56\pm 0.15$ 
and $b\pm \sigma(b)=-3.41\pm 3.22$, giving $\chi^2=6.54$.
On the other \lq physical\rq side, the energy and angular momentum dissipation
caused by dynamical friction should depend on the inner galaxy phase 
space density ($\propto \rho/\sigma^3$, where $\rho$ and $\sigma$ are
the galactic mass density and velocity dispersion, respectively). 
A stronger dynamical friction causes a faster GC decay toward inner 
galactic regions where the tidal action of a massive black hole depletes 
the GC population. Were brighter galaxies also denser in the phase-space, 
the $M_{bh}-M_V$ correlation would have a $\rho/\sigma^3$ vs. $M_V$ 
counterpart.
Using data available in the literature for a set of 428 galaxies 
(the largest part coming from a combination of data available in 
the catalogue by Prugniel \& Simien (1996)) we find the distribution 
shown in the right panel of Fig. \ref{fig4} which, far from being 
conclusive, shows indeed a trend of higher central phase space density in 
brighter galaxies.  The least square fit is
$\log (\rho/\sigma^3) = a M_V + b$ with $a\pm \sigma(a)=-0.07\pm 0.019$ 
and $b\pm \sigma(b)=-1.65\pm 0.39$, giving $\chi^2=194.44$.

\begin{table*}[!h]
\small
\begin{center}
\begin{tabular}{l c c c c} 
\hline 
{Galaxy} & {$M_V$}  & {$M_{bh}$} & {$M_l$}& {Sources} \\
\hline

\textbf{MW}& $-20.60$ & $3.61\times 10^6$& $1.80\times 10^7$ &E05, VV00, CDV97  \\

\textbf{M 31} & $-19.82$& $6.19\times 10^7$ & $2.30\times 10^7$ &M98 , CDV97  \\

\textbf{M 87} & $-22.38$ & $3.61\times 10^9$  &$2.33\times 10^9$&M98, CDV97 \\

\textbf{NGC 1427}  & $-20.43$& $1.17\times 10^8$ &$8.86\times 10^7$&VM99, CDT99 \\

\textbf{NGC 4365} & $-22.06$& $7.08\times 10^8$& $7.48\times 10^7$ &VM99, CDT99   \\

\textbf{NGC 4494} & $-20.94$ & $4.79\times 10^8$ &$2.98\times 10^7$&VM99, CDT99   \\

\textbf{NGC 4589} & $-21.14$& $3.09\times 10^8$ & $7.58\times 10^7$&VM99, CDT99  \\

\textbf{NGC 5322} & $-21.90$ & $9.77\times 10^8$  &-$6.51\times 10^7$&VM99, CDT99   \\

\textbf{NGC 5813} & $-21.81$ & $2.82\times 10^8$& $1.03\times 10^8$&VM99, CDT99  \\

\textbf{NGC 5982} & $-21.83$& $7.94\times 10^8$ & $8.86\times 10^7$ &VM99, CDT99   \\

\textbf{NGC 7626} & $-22.34$  & $1.95\times 10^9$ & $3.59\times 10^7$ &VM99, CDT99  \\

\textbf{IC 1459} & $-21.68$ & $2.60\times 10^9$ & $1.57\times 10^8$ &Fe05, VM99, CDT99  \\

\textbf{NGC 1439} & $-20.51$ & $1.95\times 10^8$&$4.79\times 10^6$ & VM99, CDT99   \\

\textbf{NGC 1700} & $-21.65$& $4.37\times 10^9$& $3.66\times 10^6$ &VM99, CDT99   \\

\textbf{NGC 1399} & $-21.71$ & $5.22\times 10^9$  &$1.44\times 10^8$ &M98, CDD01 \\

\textbf{NGC 1400} & $-20.63$ &  $3.71\times 10^8$  & $8.45\times 10^7$ &VM99, F06, CDM \\

\textbf{NGC 1407} & $-21.86$ &  $5.55\times 10^8$ & $1.31\times 10^8$&Z07, F06, CDM \\

\textbf{NGC 4472} & $-23.10$ & $2.63\times 10^9$  & $3.94\times 10^9$ &M98, RZ04, CDM   \\

\textbf{NGC 3268} & $-22.07$& $4.68\times 10^8$ &$2.98\times 10^8$ &BC06, D03, CDM \\

\textbf{NGC 3258} & $-21.40$ & $2.14\times 10^8$& $1.83\times 10^8$&BC06, D03, CDM \\

\textbf{NGC 4374}  & $-22.62$  & $1.41\times 10^9$ & $2.34\times 10^9$&R98, GR04, CDM \\

\textbf{NGC 4406} & $-22.30$ & $1.40\times 10^8$& $1.25 \times 10^9$ &CJ93, RZ04, CDM   \\

\textbf{NGC 4636} & $-21.70$  & $3.63\times 10^8$ & $6.41 \times 10^8$ &VM99, KR94, CDM  \\

\hline

\end{tabular}
\caption{col. (1) galaxy name; col. (2), (3) and (4): the $V$ absolute magnitudes, 
the galactic central black hole masses and the mass lost  
by GCSs (both in solar masses),  respectively; col. (5): bibliographic reference sources for entries in col. 
(2), (3) and (4); CDM is the present paper, the other ackronyms are defined in the References.} 
\label{tab4}
\end{center}
\end{table*}


\section{Conclusions} \label{conclusions}
We presented the comparative discussion of radial distribution of the globular 
cluster systems and of the stars in a sample of eight elliptical galaxies observed 
by various authors.
We find that GCS distributions flatten toward the centre, showing a broad core in 
the profile, contrarily to the surrounding star field. This result agrees 
with many previous findings, indicating, indeed, that GCs are usually 
less centrally concentrated than stars of the bulge-halo. A debate is still 
open on the interpretation of this observational issue. The ``evolutionary'' interpretation 
is particularly appealing; it claims that, initially, the GCS and stellar 
profiles were similar and, later, GCS evolved to the presently flatter 
distribution due to dynamical friction and tidal interactions (Capuzzo-Dolcetta 1993). In this 
picture, the flatter central profile is due to the erosion of the inner GCS radial profile. 
Many GCs are, consequently, packed in the inner galactic region, where they influence the 
physics of the host galaxy. 
Many of the galaxies studied so far have massive black holes at their centres, whose mass 
positively correlates with our estimates of number and mass of GC lost. This is a strong hint 
to the validity of the mentioned evolutionary scenario, together with the other evident 
correlation between number and mass of GC lost and their parent galaxy 
luminosity. The evolutionary hypothesis is also supported by the positive (although statistically 
uncertain) correlation between the (rough) estimate of the galactic central phase-space 
density and integrated magnitude. At the light of these encouraging findings,
we think that much effort should be spent into deepening the observational tests of this 
astrophysical scenario.


\appendix
\section{The error on the estimates of number of lost GCs} \label{app}
Here we describe how we evaluated the errors, $\epsilon_l$, given in Table \ref{tab3}. As explained in Sect. \ref{erosion}, the number of GCs lost in the galaxies of the sample has been evaluated as the integral of the difference between the (estimated) initial and present GCS radial distributions over the radial range $[r_{min}, r_{max}]$ where the two profiles differ. 
The absolute errors on $N_l$ ($\Delta N_l$) are given by the sum of the error on $N_i$ ($\Delta N_i$) and the error on $N$ ($\Delta N$), yielding the relative error $\epsilon_l=\frac{\Delta N_l}{N_l}$ of Table \ref{tab3}.
We may estimate $\Delta N$ and $\Delta N_i$ as follows.\\
i) estimate of $\Delta N$\\
For all the galaxies the number $N(r_{min}, r_{max})$ is given by
\begin{eqnarray}
N(r_{min}, r_{max})= 2\pi\Sigma_0\int_{r_{min}}^{r_{max}}\frac{r}{\left[1+\left(\frac{r}{r_c}\right)^2\right]^{\gamma}}dr=\nonumber\\=
\left.\frac{\Sigma_0\pi(r^2+r_c^2)}{(1-\gamma)\left[1+\left(\frac{r}{r_c}\right)^2\right]^{\gamma}}\right|^{r_{max}}_{r_{min}}
\end{eqnarray}

which is a function of the parameters $\Sigma_0$, $r_c$, $\gamma$, $r_{min}$ and $r_{max}$ whose indetermination is 
\begin{eqnarray} \label{deltaN}
\Delta N&=&\left|\frac{\partial{N}}{\partial{\Sigma_0}}\right|\Delta\Sigma_0+
\left|\frac{\partial{N}}{\partial{r_c}}\right|\Delta r_c+
\left|\frac{\partial{N}}{\partial{\gamma}}\right|\Delta\gamma+\nonumber\\ &+&\left|\frac{\partial{N}}{\partial{r_{min}}}\right|\Delta r_{min}+\left|\frac{\partial{N}}{\partial{r_{max}}}\right|\Delta r_{max}
\end{eqnarray}

where:

\bge
\frac{\partial{N}}{\partial{\Sigma_0}}=2 \pi\int_{r_{min}}^{r_{max}}\frac{r}{\left[1+\left(\frac{r}{r_c}\right)^2\right]^{\gamma}}dr
=\left.\frac{\pi(r^2+r_c^2)}{(1-\gamma)\left[1+\left(\frac{r}{r_c}\right)^2\right]^{\gamma}}\right|^{r_{max}}_{r_{min}},
\ede

\begin{eqnarray}
\frac{\partial{N}}{\partial{r_c}}&=&\int_{r_{min}}^{r_{max}}\frac{4\pi\Sigma_0 \gamma r^3}{r_c^3\left[1+\left(\frac{r}{r_c}\right)^2\right]^{(1+\gamma)}}dr
=\nonumber\\&=&-\left.\frac{2\pi\Sigma_0(r_c^2+\gamma r^2)}{r_c(\gamma-1)\left[1+\left(\frac{r}{r_c}\right)^2\right]^{\gamma}}\right|^{r_{max}}_{r_{min}},
\end{eqnarray}

{\setlength\arraycolsep{2pt}
\begin{eqnarray}
\frac{\partial{N}}{\partial{\gamma}}=-2 \pi \Sigma_0 \int_{r_{min}}^{r_{max}}r\left[1+\left(\frac{r}{r_c}\right)^2\right]^{-\gamma}\ln\left[1+\left(\frac{r}{r_c}\right)^2\right]dr={}
\nonumber\\
=\left.\frac{\pi \Sigma_0 (r^2+r_c^2)\left\{1+(\gamma-1)\ln\left[1+\left(\frac{r}{r_c}\right)^2\right]\right\}}{(\gamma-1)^2\left[1+\left(\frac{r}{r_c}\right)^2\right]^{\gamma}}\right|^{r_{max}}_{r_{min}},
\end{eqnarray}}

\bge\label{eq:corermin}
\frac{\partial{N}}{\partial{r_{min}}}=-2\pi\Sigma_0
\frac{r_{min}}{\left[1+\left(\frac{r_{min}}{r_c}\right)^2\right]^{\gamma}},
\ede
(we set $r_{min}=0.1$\,arcmin).
\bge
\frac{\partial{N}}{\partial{{r_{max}}}}=
2\pi\Sigma_0\frac{r_{max}}
{\left[1+\left(\frac{r_{max}}{r_c}\right)^2\right]^{\gamma}}.
\ede
The fitting parameters used to calculate $\Delta N$ are summarized in Tab. \ref{tab1}.

ii) estimate of $\Delta N_i$

The fitting formulas to the initial distribution of GCs change for the various galaxies studied.\\
For NGC 1400, NGC 1407, NGC 4374, NGC 4636 we have (see Sect. \ref{NGC 1400}, \ref{NGC 1407}, 
\ref{NGC 4374}, \ref{NGC 4636} and Tab. \ref{tab2} for the meaning and the values of the parameters)

\begin{eqnarray}\label{NiLDP}
N_i(r_{min}, r_{max})&=&2\pi\eta r_b^{-\alpha}\int_{r_{min}}^{r_{b}} r dr+2\pi\eta\int_{r_{b}}^{r_{max}}  r^{1-\alpha} dr=\nonumber\\
&=&\left.
\pi \eta r_b^{-\alpha}r^2\right|^{r_{b}}_{r_{min}}+
\left.2\pi\eta \frac{r^{2-\alpha}}{2-\alpha}\right|^{r_{max}}_{r_{b}}.
\end{eqnarray}
In Eq. \ref{NiLDP}, $\eta$ represents the parameter obtained by the vertical shifting of the luminosity profile.

The error $\Delta N_i$ is thus given by:\\
\bge
\Delta N_i= \left|\frac{\partial{N_i}}{\partial{\eta }}\right|\Delta \eta +
\left|\frac{\partial{N_i}}{\partial{\alpha}}\right|\Delta \alpha+ \left|\frac{\partial{N_i}}{\partial{{r_{min}}}}\right|\Delta r_{min} +
\left|\frac{\partial{N_i}}{\partial{{r_{max}}}}\right|\Delta r_{max}
\ede

where:

\begin{eqnarray}
\frac{\partial{N_i}}{\partial{\eta }}&=& 2\pi r_b^{-\alpha}\int_{r_{min}}^{r_{b}} r dr+2 \pi\int_{r_{b}}^{r_{max}}  r^{1-\alpha} dr=\nonumber\\
&=&\left.
\pi r_b^{-\alpha}r^2\right|^{r_{b}}_{r_{min}}+
\left.2\pi\frac{r^{2-\alpha}}{2-\alpha}\right|^{r_{max}}_{r_{b}},
\end{eqnarray}

\begin{equation}
\frac{\partial{N_i}}{\partial{r_b}}=-2\pi\eta\alpha  r_b^{-1-\alpha}\int_{r_{min}}^{r_{b}} r dr=
\left.-\pi\eta  r_b^{-1-\alpha}r^2\right|^{r_{b}}_{r_{min}},
\end{equation}

\begin{eqnarray}
\frac{\partial{N_i}}{\partial{\alpha}}&=&-2\pi\eta r_b^{-\alpha}\int_{r_{min}}^{r_{b}} r\ln(r_b) dr-2\pi\eta \int_{r_{b}}^{r_{max}} r^{1-\alpha}\ln(r) dr=\nonumber\\&=&
\left.-2\pi r_b^{-\alpha}r^2\ln(r_b)\right|^{r_{b}}_{r_{min}}+\nonumber\\ &+&
\left.\frac{2\pi\eta  r^{2-\alpha}\left\{1 + [ \alpha-2]\ln(r)\right\}}{(\gamma-2)^2}\right|^{r_{max}}_{r_{b}},
\end{eqnarray}
Also in this case we assumed $r_{min}=0.1$\,arcmin. For all the galaxies analyzed, $r_{min}>r_b$ and $r_{max}>r_b$; so we have
\begin{equation}\label{eq:rmin}
\frac{\partial{N_i}}{\partial{r_{min}}}=-2\pi\eta  r_{min}^{1-\alpha},
\end{equation}

\begin{equation}\label{eq:rmax}
\frac{\partial{N_i}}{\partial{r_{max}}}=2\pi\eta  r_{max}^{1-\alpha}.
\end{equation}

For M 49 and NGC 4406 (Sect. \ref{M 49} and Sect. \ref{NGC 4406}) we have 
\begin{eqnarray}
&&N_i(r_{min}, r_{max})=\nonumber \\&=&2\pi \eta \int_{r_{min}}^{r_{max}} r \left\lbrace{\left(\frac{r_b}{r}\right)}^{\gamma} \theta (r_b-r)+\textrm{e}^{b_n\left[\left(\frac{r_b}{r_e}\right)^{\frac{1}{n}}
-\left(\frac{r}{r_e}\right)^{\frac{1}{n}}\right]}\theta(r-r_b)\right\rbrace dr=
\nonumber\\
&=&\left.\frac{2\pi\eta}{2-\gamma}r^2
\left(\frac{r_b}{r}\right)^\gamma\right|^{r_b}_{r_{min}}+2 \pi \eta \int_{r_b}^{r_{max}} r 
\textrm{e}^{b_n\left[\left(\frac{r_b}{r_e}\right)^{\frac{1}{n}}
-\left(\frac{r}{r_e}\right)^{\frac{1}{n}}
\right]}dr
\label{eq:M49N_i}
\end{eqnarray} 
where $b_n=1.992 n-0.3271$. The second row of the previous expression is justified by the fact that, both for M 49 and NGC 4406, $r_b>r_{min}$.
Thus the error on $N_l$ 
\begin{eqnarray}
\Delta N_i&=&\left|\frac{\partial N_i}{\partial \eta}\right|\Delta \eta+ \left|\frac{\partial N_i}{\partial b_n}\right|\Delta b_n+ \left|\frac{\partial N_i}{\partial r_b}\right|\Delta r_b+ \left|\frac{\partial N_i}{\partial r_e}\right|\Delta r_e+\nonumber\\
&&+ \left|\frac{\partial N_i}{\partial n}\right|\Delta n +
\left|\frac{\partial{N_i}}{\partial{{r_{min}}}}\right|\Delta r_{min} +
\left|\frac{\partial{N_i}}{\partial{{r_{max}}}}\right|\Delta r_{max}
\end{eqnarray}
where $\Delta b_n=\left|\frac{\partial b_n }{\partial n}\right|\Delta n$, is evaluated by the following expressions of the individual error contribution:
\begin{eqnarray}\label{eq:etaM49}
\frac{\partial{N_i}}{\partial\eta}&=&
2\pi\int_{r_{min}}^{r_b}r\left(\frac{r_b}{r}\right)^\gamma dr+
2 \pi \int_{r_b}^{r_{max}} r \textrm{e}^{b_n\left[\left(\frac{r_b}{r_e}\right)^{\frac{1}{n}}
-\left(\frac{r}{r_e}\right)^{\frac{1}{n}}
\right]}dr=
\nonumber\\&=&
\left.\frac{2\pi}{2-\gamma}r^2\left(\frac{r_b}{r}\right)^\gamma\right|^{r_b}_{r_{min}}+
2 \pi \int_{r_b}^{r_{max}} r \textrm{e}^{b_n\left[\left(\frac{r_b}{r_e}\right)^{\frac{1}{n}}
-\left(\frac{r}{r_e}\right)^{\frac{1}{n}}
\right]}dr,
\end{eqnarray}
\begin{equation}
\frac{\partial{N_i}}{\partial b_n}=2 \pi \eta \int_{r_b}^{r_{max}} r \textrm{e}^{b_n\left[\left(\frac{r_b}{r_e}\right)^{\frac{1}{n}}
-\left(\frac{r}{r_e}\right)^{\frac{1}{n}}
\right]}
\left[\left(\frac{r_b}{r_e}\right)^{\frac{1}{n}}-
\left(\frac{r}{r_e}\right)^{\frac{1}{n}}\right] dr, 
\end{equation}
\begin{eqnarray}\label{eq:r_bM49}
\frac{\partial{N_i}}{\partial r_b}&=&2\pi \eta \gamma \int_{r_{min}}^{r_b}\left(\frac{r_b}{r}\right)^{\gamma-1}dr+\nonumber\\&+&2 \pi \frac{\eta b_n}{n r_e} \int_{r_b}^{r_{max}} r
\textrm{e}^{b_n\left[\left(\frac{r_b}{r_e}\right)^{\frac{1}{n}}
-\left(\frac{r}{r_e}\right)^{\frac{1}{n}}
\right]}
\left(\frac{r_b}{r_e}\right)^{\frac{1}{n}-1}dr=\nonumber\\&=&
\left.\frac{2\pi\eta \gamma}{2-\gamma}r\left(\frac{r_b}{r}\right)^{\gamma-1}\right|^{r_b}_{r_{min}}+\nonumber\\&+&2 \pi \frac{\eta b_n}{n r_e} \int_{r_b}^{r_{max}} r
\textrm{e}^{b_n\left[\left(\frac{r_b}{r_e}\right)^{\frac{1}{n}}
-\left(\frac{r}{r_e}\right)^{\frac{1}{n}}
\right]}\left(\frac{r_b}{r_e}\right)^{\frac{1}{n}-1}dr,
\end{eqnarray}
\begin{eqnarray}\\
\frac{\partial N_i}{\partial \gamma}&=&2\pi \eta \int_{r_{min}}^{r_b}r\left(\frac{r_b}{r}\right)^\gamma\ln\left(\frac{r_b}{r}\right)dr=
\nonumber\\
&=&\left.\frac{2\pi \eta r^2}{(\gamma-2)^2}\left(\frac{r_b}{r}\right)^{\gamma}\left[1+(2-\gamma)
\ln\left(\frac{r_b}{r}\right)\right]\right|^{r_b}_{r_{min}}
\end{eqnarray}
\begin{eqnarray}
\frac{\partial{N_i}}{\partial r_e}&=&-2 \pi \frac{\eta b_n}{n r_e^2} \int_{r_b}^{r_{max}} r\textrm{e}^{b_n\left[\left(\frac{r_b}{r_e}\right)^{\frac{1}{n}}
-\left(\frac{r}{r_e}\right)^{\frac{1}{n}}
\right]}\times\nonumber\\
&\times&\left[r_b\left(\frac{r_b}{r_e}\right)^{\frac{1}{n}-1}-r
\left(\frac{r}{r_e}\right)^{\frac{1}{n}-1}\right]dr,
\end{eqnarray}
\begin{eqnarray}
\frac{\partial{N_i}}{\partial n}&=&-2 \pi \frac{\eta b_n}{n^2} \int_{r_b}^{r_{max}} r\textrm{e}^{b_n\left[\left(\frac{r_b}{r_e}\right)^{\frac{1}{n}}
-\left(\frac{r}{r_e}\right)^{\frac{1}{n}}
\right]}\times\nonumber\\
&&\times\left[\left(\frac{r_b}{r_e}\right)^{\frac{1}{n}}\ln\left(\frac{r_b}{r_e}\right)-
\left(\frac{r}{r_e}\right)^{\frac{1}{n}}\ln\left(\frac{r}{r_e}\right)\right]dr,
\end{eqnarray}
Remembering that, for both NGC 4472 and NGC 4406, $r_{min}>r_b$ and $r_{max}>r_b$ we can estimate the following contributions
\begin{equation}
\frac{\partial{N_i}}{\partial{{r_{min}}}}=-2\pi \eta r_{min}\exp\left\{b_n\left[\left(\frac{r_b}{r_e}\right)^{\frac{1}{n}} -\left(\frac{r_{min}}{r_e}\right)^{\frac{1}{n}}\right]\right\},
\end{equation}
\begin{equation}\label{M49rmax}
\frac{\partial{N_i}}{\partial{{r_{max}}}}=2\pi \eta r_{max}  \exp\left\{b_n\left[\left(\frac{r_b}{r_e}\right)^{\frac{1}{n}}  -\left(\frac{r_{max}}{r_e}\right)^{\frac{1}{n}}\right]\right\}.
\end{equation}
See Tab. \ref{tab2} for the values of the parameters used in the Eq. \ref{eq:M49N_i}- Eq.\ref{M49rmax}.

Last for NGC 3268 and NGC 3258 (see Sect. \ref{NGC 3268},  Sect. \ref{NGC 3258} and Tab. \ref{tab2} for the values of the parameters) we have
\begin{equation}
\label{3268Ni}
N_i(r_{min}, r_{max})=2\pi \eta \int_{r_{min}}^{r_{max}}r\left( \frac{r_b}{r}\right)^{\gamma} 
\left[1+\left(\frac{r}{r_b}\right)^{\alpha}\right]^{\frac{\gamma-\beta}{\alpha}} dr,
\end{equation}
thus the error
\begin{eqnarray}
\Delta N_i&=&\left|\frac{\partial N_i}{\partial \eta}\right|\Delta \eta+ \left|\frac{\partial N_i}{\partial r_b}\right|\Delta r_b+ \left|\frac{\partial N_i}{\partial \gamma}\right|\Delta \gamma+\left|\frac{\partial N_i}{\partial \alpha}\right|\Delta \alpha\nonumber\\
&&+\left|\frac{\partial N_i}{\partial \beta}\right|\Delta \beta+\left|\frac{\partial{N_i}}{\partial{{r_{min}}}}\right|\Delta r_{min} +
\left|\frac{\partial{N_i}}{\partial{{r_{max}}}}\right|\Delta r_{max}
\end{eqnarray}
with
\begin{equation}
\frac{\partial{N_i}}{\partial \eta}=2 \pi \int_{r_{min}}^{r_{max}} r \left(\frac{r_b}{r}\right)^{\gamma}
\left[1+\left(\frac{r}{r_b}\right)^{\alpha}\right]^{\frac{\gamma-\beta}{\alpha}} dr,
\end{equation}
\begin{eqnarray}
 \frac{\partial{N_i}}{\partial r_b}&=&2 \pi \eta \int_{r_{min}}^{r_{max}}
\left[1+\left(\frac{r}{r_b}\right)^{\alpha}\right]^{\frac{\gamma-\beta}{\alpha}}
 \left\{\gamma \left(\frac{r_b}{r}\right)^{\gamma-1}\right.+\\ \nonumber
&
-&\left.\frac{(\gamma-\beta)}{r_b^2}r^2\left(\frac{r_b}{r}\right)^{\gamma-\alpha+1}
\left[1+\left(\frac{r}{r_b}\right)^{\alpha}\right]^{-1}\right\}dr,
\end{eqnarray}
\begin{eqnarray}
\frac{\partial{N_i}}{\partial \gamma}&=& 2 \pi \eta \int_{r_{min}}^{r_{max}}r \left(\frac{r_b}{r}\right)^{\gamma} \left[1+\left(\frac{r}{r_b}\right)^{\alpha}\right]^{\frac{\gamma-\beta}{\alpha}}\times\nonumber\\&\times&
\left\{\ln\left(\frac{r_b}{r}\right)+\frac{1}{\alpha}
\ln\left[1+\left(\frac{r}{r_b}\right)^{\alpha}\right]\right\}dr,
\end{eqnarray}
\begin{eqnarray}
\frac{\partial{N_i}}{\partial \alpha}&=& 2 \pi \eta \alpha^{-1}(\gamma-\beta) \int_{r_{min}}^{r_{max}}r\left(\frac{r_b}{r}\right)^{\gamma} \left[1+\left(\frac{r}{r_b}\right)^{\alpha}\right]^{\frac{\gamma-\beta}{\alpha}}
\times\nonumber\\&&\times
\left\{\frac{\left(\frac{r}{r_b}\right)^{\alpha} \ln\left(\frac{r}{r_b}\right)}{\left[1+\left(\frac{r}{r_b}\right)^{\alpha}\right]}
-\alpha^{-1}\ln\left[1+\left(\frac{r}{r_b}\right)^{\alpha}\right]\right\}dr,
\end{eqnarray}
\begin{eqnarray}\label{eq:3268beta}
\frac{\partial{N_i}}{\partial \beta}&=& 2 \pi \eta \alpha^{-1} \int_{r_{min}}^{r_{max}}r\left(\frac{r_b}{r}\right)^{\gamma} \left[1+\left(\frac{r}{r_b}\right)^{\alpha}\right]^{\frac{\gamma-\beta}{\alpha}}
\times\nonumber\\&\times&\ln\left[1+\left(\frac{r}{r_b}\right)^{\alpha}\right] dr.
\end{eqnarray}

\bge
\frac{\partial{N_i}}{\partial{{r_{min}}}}=-2\pi \eta r_{min}\left( \frac{r_b}{r_{min}}\right)^{\gamma} 
\left[1+\left(\frac{r_{min}}{r_b}\right)^{\alpha}\right]
^{\frac{\gamma-\beta}{\alpha}}.
\ede
with $r_{min}=0.1$\,arcmin, and
\bge
\frac{\partial{N_i}}{\partial{{r_{max}}}}=2\pi \eta r_{max}\left( \frac{r_b}{r_{max}}\right)^{\gamma} 
\left[1+\left(\frac{r_{max}}{r_b}\right)^{\alpha}\right]^{\frac{\gamma-\beta}{\alpha}}.
\ede

All the integrals from Eq. \ref{eq:M49N_i} to Eq. \ref{eq:3268beta} must be calculated 
numerically using the values of the parameters given in Table \ref{tab2}. 
The results listed in Table \ref{tab1} are obtained assuming an error 
of 1\% on each independent parameter used. Only in the case of NGC 3258 
we assumed $\Delta \gamma= 0.001$ because Capetti \& Balmaverde (2006) 
obtained $\gamma=0$ from their Nuker fit, and so it is impossible to 
give an estimate of the error as a percentage of  $\gamma$.

\begin{acknowledgements}
      
\end{acknowledgements}

\bibliographystyle{aa} 
\end{document}